\documentclass{article}
\usepackage{pstricks}
\usepackage{latexsym}
\usepackage{amssymb}

\newcommand{\textlineskip}{\baselineskip=13pt}

\def\qed{\hbox{${\vcenter{\vbox{\hrule height 0.4pt\hbox{\vrule width 0.4pt height 6pt
\kern5pt\vrule width 0.4pt}\hrule height 0.4pt}}}$}}
\begin{document}

\centerline{\bf
A NOTE ON QUANTUM ONE-WAY PERMUTATIONS}

\vspace*{0.37truein}
\centerline{\footnotesize ELHAM KASHEFI}
\vspace*{0.015truein}
\centerline{\footnotesize\it Centre for Quantum Computation, Clarendon 
Laboratory, University of Oxford, Parks Road}
\baselineskip=10pt
\centerline{\footnotesize\it Oxford OX1 3PU, England}
\baselineskip=10pt
\centerline{\footnotesize\it Optics Section, The Blackett Laboratory,
Imperial College}
\baselineskip=10pt
\centerline{\footnotesize\it London SW7 2BZ, England}
\vspace*{10pt}
\centerline{\footnotesize HARUMICHI NISHIMURA}
\vspace*{0.015truein}
\centerline{\footnotesize\it Centre for Quantum Computation, Clarendon 
Laboratory, University of Oxford, Parks Road}
\baselineskip=10pt
\centerline{\footnotesize\it Oxford OX1 3PU, England}
\vspace*{10pt}
\centerline{\footnotesize VLATKO VEDRAL}
\vspace*{0.015truein}
\baselineskip=10pt
\centerline{\footnotesize\it Optics Section, The Blackett Laboratory,
Imperial College}
\baselineskip=10pt
\centerline{\footnotesize\it London SW7 2BZ, England}

\vspace*{0.21truein}
\begin {abstract}
We discuss the question of the existence of quantum one-way permutations. First, 
we prove the equivalence between inverting a permutation and that of 
constructing a polynomial size network for reflecting about a given quantum 
state. Next, we consider the question: if a state is difficult to prepare, is 
the 
operator reflecting about that state difficult to construct? By revisiting 
Grover's algorithm, we present the relationship between this question and the 
existence of one-way permutations. Moreover, we compare our method to Grover's 
algorithm and discuss possible applications of our results.
\end {abstract}

\vspace*{1pt}\textlineskip
\section{Introduction}
\vspace*{-0.5pt}
\noindent
Quantum computation is a rapidly growing field which explores the relationship 
between quantum physics and computation \cite{NC00}. We have two strong 
indications that quantum systems are potentially more efficient than their 
classical counter-parts at performing  computational tasks. One is Shor's 
algorithm 
\cite{Shor94}, which solves the factoring problem and the discrete logarithm 
problem in quantum polynomial time. The other is Grover's algorithm 
\cite{Grover96}, which works quadratically faster than any classical algorithm 
for the search problem in the oracle setting. On the other hand, Bennett, 
Bernstein, Brassard, and Vazirani \cite{BBBV97} have shown that with probability 
$1$ there exists a quantum one-way permutation relative to a random permutation 
oracle.

The existence of one-way functions is one of the most important open problems in 
classical computation. For example, it is well-known that one-way functions have 
applications in cryptography \cite{Papadimi94}. Loosely speaking, a one-way 
function is one that is easy to compute but hard to invert. To make this notion 
precise, we define a function $f$ to be {\it (quantum) one-way}, if $f$ is 
one-one, $f$ is honest, $f$ can be computed in (quantum) polynomial time, and 
$f^{-1}$ is not computable in (quantum) polynomial time. By $f$ being honest we 
mean that there exists a polynomial $p$ such that $|x|\leq p(|f(x)|)$, where 
$|.|$ denotes the length of binary strings. Note that in this paper we are 
discussing one-way functions in the setting of the worst case complexity \cite 
{GS84}. It is thought that the proof of existence of one-way function is a 
difficult problem, since it is equivalent to the separation between the 
complexity classes {\bf P} and {\bf UP} \cite {GS84}. 

We address the question of the existence of quantum one-way permutations which 
is a restricted type of one-way functions. First, we consider a necessary and 
sufficient condition for inverting efficiently a polynomial time computable 
permutation. In the classical case, Hemaspaandra and Rothe \cite{HR00} presented 
a 
necessary and sufficient condition for the existence of one-way permutations. We 
show that in the quantum setting, the problem of inverting a permutation in 
polynomial time is equivalent to the problem of constructing polynomial size 
networks for the reflections about some quantum states $|\psi\rangle$, i.e., 
$2|\psi\rangle\langle\psi|-I$. In the proof of this equivalence, we present a 
quantum algorithm for inverting a permutation efficiently under the condition 
that the reflections about their quantum states are efficiently implementable. 
Similar to Grover's algorithm, our algorithm also consists of the iteration of 
the tagging and reflection operators \cite{Grover96}. We show that the 
exponential speed-up over Grover's algorithm is possible if and only if the 
efficient reflections about some quantum states that we will define in the 
paper, are possible. 

Next, we consider the relationship between the complexity of preparing a state 
and the reflection about that state. We define a {\it unitary operator} on $n$ 
qubits to be {\it easy} if there exists a polynomial size network implementing 
the operator up to a global phase. The $n$-qubit {\it state} $|\phi\rangle$ is 
defined to be {\it easy} if there exists a polynomial size network which produces the state $|\phi\rangle$ up to a global phase. 
It is straightforward to see that if a state is easy, the reflection about that 
state is also easy. We consider the other direction, which seems to hold at 
first glance. {\it If the reflection about a state is easy, the state itself is 
easy} (we will refer to this statement as Assumption A). However, by exposing 
another view of Grover's algorithm, we can find a counter-example such that 
Assumption A is false if a quantum one-way permutation exists.

This paper is organized as follows. In Section 2, we prove the equivalence 
between inverting a permutation and constructing a quantum network implementing 
the 
reflections about some special quantum states. In Section 3, we revisit Grover's 
algorithm from the viewpoint of Assumption A, and show that Assumption A is 
false if a quantum one-way permutation exists. Moreover, we compare our 
algorithm with Grover's algorithm in the light of Assumption A. In Section 4, we 
discuss other related results and possible applications of our results.   

\section{Main Result}
\noindent

For any permutation $f$ on $n$-bit strings, let $U_f$ denote the unitary 
operator mapping the basis state $|x\rangle|y\rangle$ to $|x\rangle|f(x)\oplus 
y\rangle$, where $|x\rangle$ and $|y\rangle$ each consist of $n$ qubits. We 
consider the following problem called hereafter {\bf INVERT}: for any given 
$x\in\{0,1\}^n$, find $f^{-1}(x)$. In the setting where $f$ is given as an 
oracle, Grover's algorithm can solve {\bf INVERT} with quadratic speed-up over 
any 
classical algorithm \cite{Grover96}. In his algorithm, Grover uses the tagging 
operator $O$ defined as
$$
O|x\rangle|y\rangle=(-1)^{\delta_{x,f(y)}}|x\rangle|y\rangle
$$
and the reflection about the uniform superposition defined as
$$
|\psi\rangle=\sum_{y\in\{0,1\}^n}|y\rangle,
$$
i.e. $2|\psi\rangle \langle\psi| - I$, which is also called the inversion about 
the average amplitude. The operator $O$ can be simulated by two applications 
of $U_f$ and $n$ controlled-not gates. Moreover, if $f$ is polynomial time computable, then it is also possible to efficiently construct the unitary 
operator 
$O[k]$ defined by 
$$
O[k]|x\rangle|y\rangle=(-1)^{\delta_{x_k,f_k(y)}\cdot\delta_{x_{k+1},f_{k+1}(y)}
}|x\rangle|y\rangle,
$$
where $x_i$ and $f_i(y)$ for $i=1,\ldots,n$ represent the $i$-th bits of $x$ and 
$f(y)$. This operator $O[k]$ will enable us to mark all the states $|y\rangle$ 
such that 2 qubits of $|f(y)\rangle$ are equal to the corresponding qubits of 
$|x\rangle$. Geometrically, $O[k]$ can be considered to be the reflection about 
the hyper-plane spanned by the vectors $\{|y\rangle|\  f(y)_{(k,k+1)}\neq 
x_{(k,k+1)}\}$. We show that if we can efficiently implement $O[k]$'s and the 
unitary operator 
$$
Q_j=\sum_{x\in\{0,1\}^n}|x\rangle\langle x|\otimes 
(2|\psi_{j,x}\rangle\langle\psi_{j,x}|-I),
$$
where
$$
|\psi_{j,x}\rangle=\frac{1}{\sqrt{2^{n-2j}}} 
\sum_{y:f(y)_{(1,2j)}=x_{(1,2j)}}|y\rangle,  
$$
then we can efficiently invert $f$ by polynomial size network. Conversely, we 
also can show that if $f$ is difficult to invert, then $Q_j$'s are also 
difficult to construct. 

Now we state and prove this result formally. We say that a set $F$ of unitary 
operators is {\it easy} if every $U \in F$ is easy. 

\vspace*{12pt}
\noindent
{\bf Theorem~1:} A function $f:\{0,1\}^n\rightarrow\{0,1\}^n$ is a quantum 
one-way permutation if and only if the set 
$F_n=\{Q_j\}_{j=0,1,\ldots,\frac{n}{2}-1}$ of unitary operators is not easy.

\vspace*{12pt}
\noindent
{\bf Proof:} Without loss of generality, we can assume that $n$ is even. 

($\Rightarrow$) Suppose that $F_n$ is easy. Then we show that $f^{-1}$ is 
computable by a polynomial size quantum network. A quantum algorithm (Algorithm 
A 
below) computing $f^{-1}$ is as follows. Assume that $x$ is given as the input 
in the first register of the quantum network to be constructed.

\

ALGORITHM A

\vskip5pt
Step 1 (Preparation). 

Prepare the second register in the uniform superposition
$$
|\psi_0\rangle=\frac{1}{\sqrt{2^n}}\sum_{y\in\{0,1\}^n}|y\rangle.
$$

Step 2 (Iteration). 

For $j=0$ to $\frac{n}{2}-1$, implement the following steps 2.j.1--2.j.2.

Step 2.j.1. Carry out $O[2j+1]$ on the first and the second registers. 

Step 2.j.2. Carry out $Q_j$ on the first and the second registers.

\

Step 2.j.1 can be implemented through the following 3 steps: (1) Carry out 
$U_f:|y\rangle|z\rangle\mapsto|y\rangle|f(y)\oplus z\rangle$ on the second and 
third registers. (2) Compare the $2j+1$-th and the $2j+2$-th qubits of the first 
register with the corresponding qubits of the third register, and apply a phase 
shift of $-1$ if they are same; otherwise do nothing. (3) Carry out $U_f$ on the 
second and third registers. 

Now we show that Algorithm A computes $f^{-1}$. After Step 1, the state of the 
system is 
$$
\frac{1}{\sqrt{2^n}}|x\rangle\sum_{y\in\{0,1\}^n}|y\rangle.
$$
We show that after Step 2.j.2 the state of the system is
$$
\frac{2^{j+1}}{\sqrt{2^n}}|x\rangle \sum_{y:f(y)_{(1,2j+2)}=x_{(1,2j+2)}}|y \rangle \, ,
$$
which means that Algorithm A computes $f^{-1}$ after $\frac{n}{2}$ iterations. 
In the case $j=0$, the state evolves as follows (note that for any $x$ we 
have $|\psi_{0,x}\rangle=|\psi_0\rangle$)
\begin{eqnarray*}
& &\frac{1}{\sqrt{2^n}}|x\rangle\sum_{y\in\{0,1\}^n}|y\rangle\nonumber\\
&\stackrel {\mbox {\scriptsize $2.0.1$}}{\longrightarrow}& 
\frac{1}{\sqrt{2^n}}|x\rangle \left(\sum_{y:f(y)_{(1,2)}\neq x_{(1,2)}}|y\rangle 
-\sum_{y:f(y)_{(1,2)}=x_{(1,2)}}|y\rangle\right)\nonumber\\
& & =\frac{1}{\sqrt{2^n}}|x\rangle \left(\sqrt{2^n}|\psi_0\rangle 
-2\sum_{y:f(y)_{(1,2)}=x_{(1,2)}}|y\rangle\right)\nonumber\\
&\stackrel {\mbox {\scriptsize $2.0.2$}}{\longrightarrow}& 
\frac{1}{\sqrt{2^n}}|x\rangle (2|\psi_0\rangle\langle\psi_0|-I) 
\left(\sqrt{2^n}|\psi_0\rangle 
-2\sum_{y:f(y)_{(1,2)}=x_{(1,2)}}|y\rangle\right)\nonumber\\
& &=\frac{1}{\sqrt{2^n}}|x\rangle \left( 
2\sqrt{2^n}|\psi_0\rangle-\sqrt{2^n}|\psi_0\rangle 
-4|\psi_0\rangle\sum_{y:f(y)_{(1,2)}=x_{(1,2)}}\langle\psi_0|y\rangle\right) 
\nonumber\\
& & +2\!\!\sum_{y:f(y)_{(1,2)}=x_{(1,2)}}|y\rangle\nonumber\\
& &=\frac{2}{\sqrt{2^n}}|x\rangle \sum_{y:f(y)_{(1,2)}=x_{(1,2)}}|y\rangle.
\end{eqnarray*}
On the other hand, suppose that the case $j=k-1$ holds. Then, following Steps 
2.k.1--2.k.2, the state evolves as follows
\begin{eqnarray*}
& &\frac{2^k}{\sqrt{2^n}}|x\rangle \sum_{y:f(y)_{(1,2k)}=x_{(1,2k)} 
}|y\rangle\nonumber\\
&\stackrel {\mbox {\scriptsize 
$2.k.1$}}{\longrightarrow}&\frac{2^k}{\sqrt{2^n}}|x\rangle\left(\sum_{y:f(y)_{(1
,2k)}=x_{(1,2k)}}|y\rangle 
-2\sum_{y:f(y)_{(1,2k+2)}=x_{(1,2k+2)}}|y\rangle\right)\nonumber\\
& &=\frac{2^k}{\sqrt{2^n}}|x\rangle 
\left(\sqrt{2^{n-2k}}|\psi_{k,x}\rangle-2\sum_{y:f(y)_{(1,2k+2)}=x_{(1,2k+2)}}|y
\rangle\right)\nonumber\\
&\stackrel {\mbox {\scriptsize $2.k.2$}}{\longrightarrow}& 
\frac{2^k}{\sqrt{2^n}}|x\rangle (2|\psi_{k,x}\rangle\langle\psi_{k,x}|-I) 
\left(\sqrt{2^{n-2k}}|\psi_{k,x}\rangle-2\sum_{y:f(y)_{(1,2k+2)}=x_{(1,2k+2)}}|y
\rangle\right)\nonumber\\
& &=\frac{2^k}{\sqrt{2^n}}|x\rangle 
\left(2\sqrt{2^{n-2k}}|\psi_{k,x}\rangle-\sqrt{2^{n-2k}}|\psi_{k,x}\rangle 
-4|\psi_{k,x}\rangle\sum_{y:f(y)_{(1,2k+2)}=x_{(1,2k+2)}} 
\langle\psi_{k,x}|y\rangle\right)\nonumber\\
& 
&+\frac{2^k}{\sqrt{2^n}}|x\rangle\left(2\sum_{y:f(y)_{(1,2k+2)}=x_{(1,2k+2)}}|y \rangle \right)\nonumber\\
& 
&=\frac{2^{k+1}}{\sqrt{2^n}}|x \rangle \sum_{y:f(y)_{(1,2k+2)}=x_{(1,2k+2)}} |y \rangle \, .
\end{eqnarray*}
Thus, the case $j=k$ holds. From the assumption that $\{ Q_j \}$ is easy, it is 
simple to see that Algorithm A can be implemented by a polynomial size quantum 
network.

\vskip5pt
($\Leftarrow$) Suppose that $f$ is not a one-way permutation. Then we show that 
$\{Q_j\}_{j=0,1,\ldots,\frac{n}{2}-1}$ can be implemented by a polynomial size 
quantum network. According to the assumption, $f$ and $f^{-1}$ are quantum 
polynomial time computable. The following operator 
$$
M_f:|x\rangle\mapsto|f(x)\rangle
$$
can be implemented by a polynomial size quantum network \cite{Bennett73,KKVB01}. 
To see why note that :
$$
M_f\otimes I = ( U_{f^{-1}} )^{-1} \otimes S \otimes U_f \, , 
$$
where the swap gate $S$ is defined by $S:|a\rangle \otimes 
|b\rangle\rightarrow|b\rangle \otimes |a\rangle$. 

We now show that the unitary operator $Q'_j=(I\otimes M_f)Q_j (I\otimes 
M_f)^\dagger$ can be implemented by a polynomial size quantum network, which 
means 
that $Q_j$ can also be implemented by a polynomial size quantum network. The 
operator $Q'_j$ can be rewritten as follows

\begin{eqnarray*}
Q'_j &=& (I\otimes M_f) \left\{\sum_{x\in\{0,1\}^n}|x\rangle\langle x|\otimes 
\left(2\left(\frac{1}{2^{n-2j}}\sum_{y,y'}{}^* |y\rangle\langle y'|\right)-I 
\right) \right\}(I\otimes M_f)^\dagger\nonumber\\
&=& \sum_{x\in\{0,1\}^n}|x\rangle\langle 
x|\otimes\left(2\frac{1}{2^{n-2j}}\sum_{y,y'}{}^* |f(y)\rangle\langle f(y')|-I 
\right)\nonumber\\ &=&  \sum_{x\in\{0,1\}^n}|x\rangle\langle 
x|\otimes \left( 2|x_{(1,2j)}\rangle\langle x_{(1,2j)}| 
\frac{1}{2^{n-2j}}\sum_{y,y'}{}^* |f(y)_{(2j+1,n)}\rangle\langle 
f(y')_{(2j+1,n)}| 
-I\right)\nonumber\\
&=& \sum_{x\in\{0,1\}^n}|x\rangle\langle x|\otimes \left( 
2|x_{(1,2j)}\rangle\langle x_{(1,2j)}| 
\otimes|\psi_j\rangle\langle\psi_j|-I\right)\nonumber\\
&=& \sum_{x\in\{0,1\}^n}|x\rangle\langle x|\otimes 
\left(|x_{(1,2j)}\rangle\langle x_{(1,2j)}| 
\otimes(2|\psi_j\rangle\langle\psi_j|-I) +\sum_{y:y\neq 
x_{(1,2j)}}|y\rangle\langle y|\otimes I \right).
\end{eqnarray*}
Here, $\sum_{y,y'}^*$ denotes 
$\sum_{y,y':f(y)_{(1,2j)}=f(y')_{(1,2j)}=x_{(1,2j)}}$ and $|\psi_j\rangle$ 
denotes
$$
|\psi_j\rangle=\frac{1}{\sqrt{2^{n-2j}}} \sum_{i\in\{0,1\}^{n-2j}}|i\rangle.
$$
Thus, we can implement $Q'_j$ by comparing the first $2j$ qubits of the first 
register with the corresponding qubits of the second register and applying 
$2|\psi_j\rangle\langle\psi_j|-I$ if they are the same and applying the identity 
otherwise (i.e. conditional-$(2|\psi_j\rangle\langle\psi_j|-I)$). The 
operator $2|\psi_j\rangle\langle\psi_j|-I$ is easy, since 
$2|\psi_j\rangle\langle\psi_j|-I=H^{\otimes n-2j}(2|0\rangle\langle 
0|-I)H^{\otimes n-2j}$, where 
$H$ is the Hadamard gate and the superscript $n-2j$ indicates that Hadamard gate 
is applied to the last $n-2j$ qubits. Therefore, $Q'_j$ is easy and this 
completes the proof.\ \ \qed\,

Note that all unitary operators $U_k$ are easy if and only if the operation 
$$
\sum_k|k\rangle\langle k|\otimes U_k,
$$
which implements $U_k$ conditionally, is easy. The operator $Q_j$ implements the 
reflection about the state $|\psi_{j,x}\rangle$ conditionally, therefore Theorem 
$1$ gives a necessary and sufficient condition for quantum one-way permutations 
in terms of the reflection about a quantum state. 

\section{General View}
\noindent

It is well-known that if a state is easy, then the reflection about the state is 
easy \cite{NC00}. Does the inverse hold? We call its inverse, i.e., the 
statement ``if the reflection about a state is easy, the state itself is easy'', 
{\bf Assumption A}. In this section, we revisit both Grover's and our algorithm 
from the viewpoint of complexity of a state and the reflection about the state, 
and discuss the relationship between the existence of one-way permutations and 
Assumption A. 
 
First, let us revisit Grover's search problem and algorithm. In \cite{Grover96}, 
Grover considered the following problem called hereafter {\bf SEARCH}. Let 
$f:\{0,1\}^n\rightarrow\{0,1\}$ be a function such that $|f^{-1}(\{1\})|=1$. 
Then, the goal is to find $f^{-1}(1)$. Grover's algorithm for this problem 
(Algorithm B below) 
consists of the following steps.

\

ALGORITHM B

\vskip5pt
Step 1 (Preparation). 

Prepare the uniform superposition 
$$
|\psi\rangle=\sum_{x\in\{0,1\}^n}|x\rangle.
$$

Step 2 (Iteration).

Iterate Step 2.1 and Step 2.2. 

Step 2.1.  
Carry out the tagging operation given by
$$
\sum_{x\in\{0,1\}^n}(-1)^{\delta_{x,f^{-1}(1)}}|x\rangle\langle 
x|=I-2|f^{-1}(1)\rangle\langle f^{-1}(1)|.
$$ 

Step 2.2.  
Carry out the reflection about the state $|\psi\rangle$ (i.e. the inversion 
about the average amplitude). 

\

Here, Step 1 and Step 2.2 are easy, and Step 2.1 can be implemented by querying 
the oracle 
$$
U_f:|x\rangle|b\rangle\rightarrow |x\rangle|f(x)\oplus b\rangle
$$
twice, since for the reflection about the target state $|f^{-1}(1)\rangle$, we 
have  
$$
\{(2|f^{-1}(1)\rangle\langle f^{-1}(1)|-I)\otimes 
I\}|x\rangle|0\rangle=\{U_f(I\otimes(2|1\rangle\langle 
1|-I))U_f\}|x\rangle|0\rangle. 
$$
By $O(\sqrt{2^n})$ iterations of Step 2, i.e., by $O(\sqrt{2^n})$ queries, we 
can get $f^{-1}(1)$ with high probability. Thus, when $f$ is given as an oracle, 
this algorithm works quadratically faster than any possible classical algorithm. 
However, this algorithm is shown to be optimal \cite{BBBV97,BBHT96,Zalka99}, 
so that we cannot solve this problem by using at most queries polynomial in $n$. 
This implies that even if the reflection about the state $|f^{-1}(1)\rangle$ 
is {\it assumed to be} easy, the state $|f^{-1}(1)\rangle$ itself is not easy. 

Next, we shall relate Assumption A to the existence of quantum one-way 
permutations by revisiting Grover's algorithm for {\bf INVERT} considered in the 
previous section. Grover's algorithm for {\bf INVERT} (Algorithm C below) is as 
follows.

\

ALGORITHM C

\vskip5pt
Step 1 (Preparation).  

Prepare the uniform superposition
$$
|\psi\rangle=\sum_{y\in\{0,1\}^n}|y\rangle.
$$

Step 2 (Iteration).  

Iterate Step 2.1 and Step 2.2. 

Step 2.1.  
Carry out the tagging operator 
$$
O=I-2|f^{-1}(x)\rangle\langle f^{-1}(x)|.
$$

Step 2.2.  
Carry out the reflection about the state $|\psi\rangle$. 

\

Similar to Algorithm A, Step 1 and Step 2.2 are easy. The operator $O$ in Step 
2.1 is a tagging operator and can be implemented by using the operator
$$
U_f:|y\rangle|z\rangle\rightarrow |y\rangle|f(y)\oplus z\rangle.
$$
In fact, for any $y\in\{0,1\}^n$ we have 
$$
\{(I-2|f^{-1}(x)\rangle\langle f^{-1}(x)|)\otimes 
I\}|y\rangle|0\rangle=U_f(I\otimes(I-2|x\rangle\langle 
x|))U_f|y\rangle|0\rangle. 
$$
Thus, given $U_f$ as an oracle, we can compute $f^{-1}(x)$ with high probability 
by $O(\sqrt{2^n})$ queries. This algorithm is also shown to be optimal 
\cite{Ambainis00}. Note that the operator $2|f^{-1}(x)\rangle\langle 
f^{-1}(x)|-I$ is performing the reflection about the state $|f^{-1}(x)\rangle$. 
Thus, 
Algorithm C shows that even if the reflection about the state 
$|f^{-1}(x)\rangle$ is {\it assumed to be} easy, the state itself is not 
necessarily easy. 

Now, let us consider the case when $f$ is a quantum one-way permutation. Then, 
the operator $U_f$ is easy, so that $2|f^{-1}(x)\rangle\langle f^{-1}(x)|-I$ is 
also easy. Therefore, from Algorithm C, we can infer the following interesting 
fact: {\it If there exists a quantum one-way permutation, there exists a 
counter-example to Assumption A}.  

Compared to Algorithm C, the Algorithm A for inverting $f$ is exponentially 
faster, providing that $Q_j$'s are easy. The operator $Q_j$ is a unitary 
operator 
which reflects any given state about the state $|\psi_{j,x}\rangle$, where $x$ 
is the value of the first register. Does the state $|\psi_{j,x}\rangle$ also 
provide a counter-example to Assumption A? The answer is no, since it is shown 
in the proof of Theorem $1$ that if $F_n$ is easy, we can generate the state 
$|\psi_{j,x}\rangle$ by a polynomial size network. Therefore we have the 
following corollary from Theorem $1$. 

\vspace*{12pt}
\noindent
{\bf Corollary~2:} The following conditions are equivalent. 
\begin{itemize}
\item $f$ is not a quantum one-way permutation.
\item All states $|\psi_{j,x}\rangle$, where $j\in\{0,\frac{n}{2}-1\}$ and 
$x\in\{0,1\}^n$, are easy. 
\item The reflections about all states $|\psi_{j,x}\rangle$ are easy. 
\end{itemize}

So far we have considered only the exact setting. However, using the diamond 
metric and its properties \cite{AKN98}, the similar results also hold in the 
bounded error setting. In the latter setting we define the notions of easy 
operator, easy state, and quantum one-way function as follows. A 
trace-preserving completely positive superoperator (CPSO) $U$ is defined to be 
{\it approximately easy} if there exists a family of polynomial size quantum 
networks $\{ N_{\epsilon}\}$ such that
$$
||U - U_{\epsilon}||{}_{\Diamond}\leq \epsilon \, ,
$$
where $U_{\epsilon}$ is the operator implemented by $N_{\epsilon}$ exactly. A 
mixed state $\rho$ is defined to be {\it approximately 
easy} if there exists an approximately easy CPSO $U$ such that 
$$
U(|0\rangle\langle 0|)= \rho \, .
$$
Now we can give the definition of quantum one-way function in the bounded-error 
setting. A function $f$ is {\it quantum one-way} if $f$ is one-one, $f$ is 
honest, $f$ is approximately easy, and $f^{-1}$ is not approximately easy, where 
$f$ is quantum approximately easy if the unitary operator $U_f: 
|x\rangle|z\rangle \mapsto |x\rangle|z\oplus f(x)\rangle$ is approximately easy. 
It is straightforward to check that if there exists a quantum one-way 
permutation, then there exists a counter-example for the Assumption A in the 
bounded error setting.

\section{Discussions}
\noindent

We have reduced the problem of the existence of a quantum one-way permutation to 
the problem of constructing a polynomial size network for performing the 
specific task of the reflection about a given state. Ambainis \cite{Ambainis00} 
proved that inverting a permutation on the $n$-bit strings in the standard query 
model requires $\Omega(\sqrt{2^n})$ queries. In the standard query model 
\cite{BBCMW98}, a quantum computation with $T$ queries is a sequence of unitary 
operators 
$$
U_0\rightarrow O\rightarrow U_1\rightarrow O\cdots\rightarrow U_{T-1}\rightarrow 
O\rightarrow U_T,
$$ 
where $U_j$'s are arbitrary unitary operators independent of the input qubits, 
and $O$ is the standard query operator. However, our algorithm is consistent 
with Ambainis' result, since we consider the case that $U_j$'s depend on the 
input qubits and this does not fit his model. 

Another related issue is the work of Chen and Diao \cite{CD00} where they 
attempted to present an efficient quantum algorithm for the problem SEARCH, 
which is similar to our algorithm for the problem INVERT. They mentioned that 
the tagging operation and the reflection about a given state which varies 
dynamically can be constructed by polynomial size networks, but they did not 
show the construction for their operations. (This construction is, of course, 
impossible given Grover's black box, since it would violate the optimality proof 
of Grover's algorithm \cite{BBHT96,Zalka99,Ambainis00}.) For our problem INVERT 
we have given the polynomial size network of the tagging operation and we have 
shown that the difficulty of the construction of the reflection operation is 
equivalent to the existence of the quantum one-way permutation. Furthermore it 
is an interesting open problem whether there exists a reduction from other types 
of one-way functions to constructing a polynomial size network for performing 
the reflection about a given state.

On the other hand, we have seen that Grover's algorithm gives us an example of 
states that are difficult to prepare but the reflections about these states are 
easy, i.e., it provides a counter-example to Assumption A assuming the existence 
of one-way permutations. This investigation of Assumption A seems to be useful 
for cryptographic applications since recently, quantum bit commitment protocols 
based on quantum one-way permutations have been proposed \cite{DMS00,AC01}. 
Moreover, it is interesting to find such a concrete counter-example without the 
existence of quantum one-way permutations. Presenting such examples of states 
may provide us with more ideas for constructing novel quantum algorithms.

\vspace*{10pt}
\noindent
{\bf Acknowledgements}
\noindent
This work was
supported by EPSRC, the European grant EQUIP and the QUIPROCONE grant.

\end{document}